\documentclass[aps,prb,twocolumn,showpacs, notitlepage]{revtex4-2}
\usepackage{graphicx}% Include figure files
\usepackage{natbib}
\usepackage[utf8]{inputenc}
\usepackage{epstopdf}
\setcitestyle{super}
\usepackage{color}
\usepackage{amsmath}
\usepackage[normalem]{ulem}
\usepackage{pdfpages}
\usepackage{hyperref}
\usepackage{tabularx}
\makeatletter
\AtBeginDocument{\let\LS@rot\@undefined}
\makeatother

\newcommand*{\citen}[1]{%
  \begingroup
    \romannumeral-`\x % remove space at the beginning of \setcitestyle
    \setcitestyle{numbers}%
    \cite{#1}%
  \endgroup   
}

\usepackage{xcolor}

\usepackage{hyphenat}
\begin{document}

\preprint{APS/123-QED}

\title{Symmetric versus antisymmetric strain tuning of the valence transition in Yb(In$_{1-x}$Ag$_x$)Cu$_4$}

\author{Caitlin I. O'Neil$^{1,2}$}
\author{Michelle Ocker$^{3}$}
\author{Kristin Kliemt$^{3}$}
\author{Cornelius Krellner$^{3}$}
\author{Elena Gati$^{1,4}$}\email{elena.gati@cpfs.mpg.de}

\address{$^{1}$ Max Planck Institute for Chemical Physics of Solids, 01187 Dresden, Germany}
\address{$^{2}$ Scottish Universities Physics Alliance, School of Physics and Astronomy, University of St Andrews, St Andrews KY16 9SS, UK}
\address{$^{3}$ Physikalisches Institut, Goethe Universität, 60439 Frankfurt, Germany}
\address{$^{4}$ Institut für Festkörper- und Materialphysik, Technische Universität Dresden, 01062 Dresden, Germany}

\date{\today}

\begin{abstract}

Similar to transitions in a range of correlated quantum materials, the valence transition exhibits a strong coupling to the crystal lattice, rendering it highly sensitive to stress tuning. In the present work, we determine the effect of uniaxial stress, which breaks the lattice symmetry, on the valence transition temperature and its crossover temperature in pure and Ag-substituted YbInCu$_4$. Our key result is that hydrostatic stress is more effective in tuning this transition than uniaxial stress. Based on a symmetry decomposition of the stress-induced strains, we argue that this observation can be quantitatively understood, given that the valence transition is mostly sensitive to symmetric strains and thus volume changes of the lattice. These results support the notion that the valence transition can give rise to \textit{critical elasticity} close to its critical endpoint.

\end{abstract}

\pacs{xxx}

\maketitle

\section{Introduction}

The strong coupling between electronic and lattice degrees of freedom in correlated quantum materials makes their electronic properties highly sensitive to external stress and strain. Likewise, the electronic system can exert a significant influence on the lattice response. Importantly, this influence may go beyond a small, perturbative response. Rather than adhering to Hooke’s law with a linear stress-strain relationship, these systems can show a strongly non-linear lattice response, signaling a non-perturbative interaction between electrons and the crystal lattice.

Striking examples of such behavior include the breakdown of Hooke’s law near the Mott transition in an organic conductor \cite{Gati16}, the pronounced lattice softening accompanying the electronic Lifshitz transition in the unconventional superconductor Sr$_2$RuO$_4$\,\cite{Noad23}, and the significant renormalization of the Young’s modulus observed at the nematic transition in iron-based superconductors \cite{Boehmer16}.

Crucially, this strong electron-lattice coupling alters the fundamental nature of phase transitions in these systems. The long-range nature of the lattice forces acts to suppress electronic fluctuations, which in turn drives such transitions toward mean-field behavior. As a result, even though these transitions are driven by electronic degrees of freedom, the lattice also becomes critical and determines the universal behavior of the electronic transition. This phenomenon, known as \textit{critical elasticity} \cite{Zacharias12, Zacharias15}, has been experimentally confirmed at the Mott finite-temperature critical endpoint \cite{Gati16} and theoretically proposed to govern the behavior at nematic \cite{Paul17}, metamagnetic \cite{Zacharias13}, and altermagnetic \cite{Steward25} quantum critical points.

\begin{figure*}
    \centering
    \includegraphics[width = \textwidth]{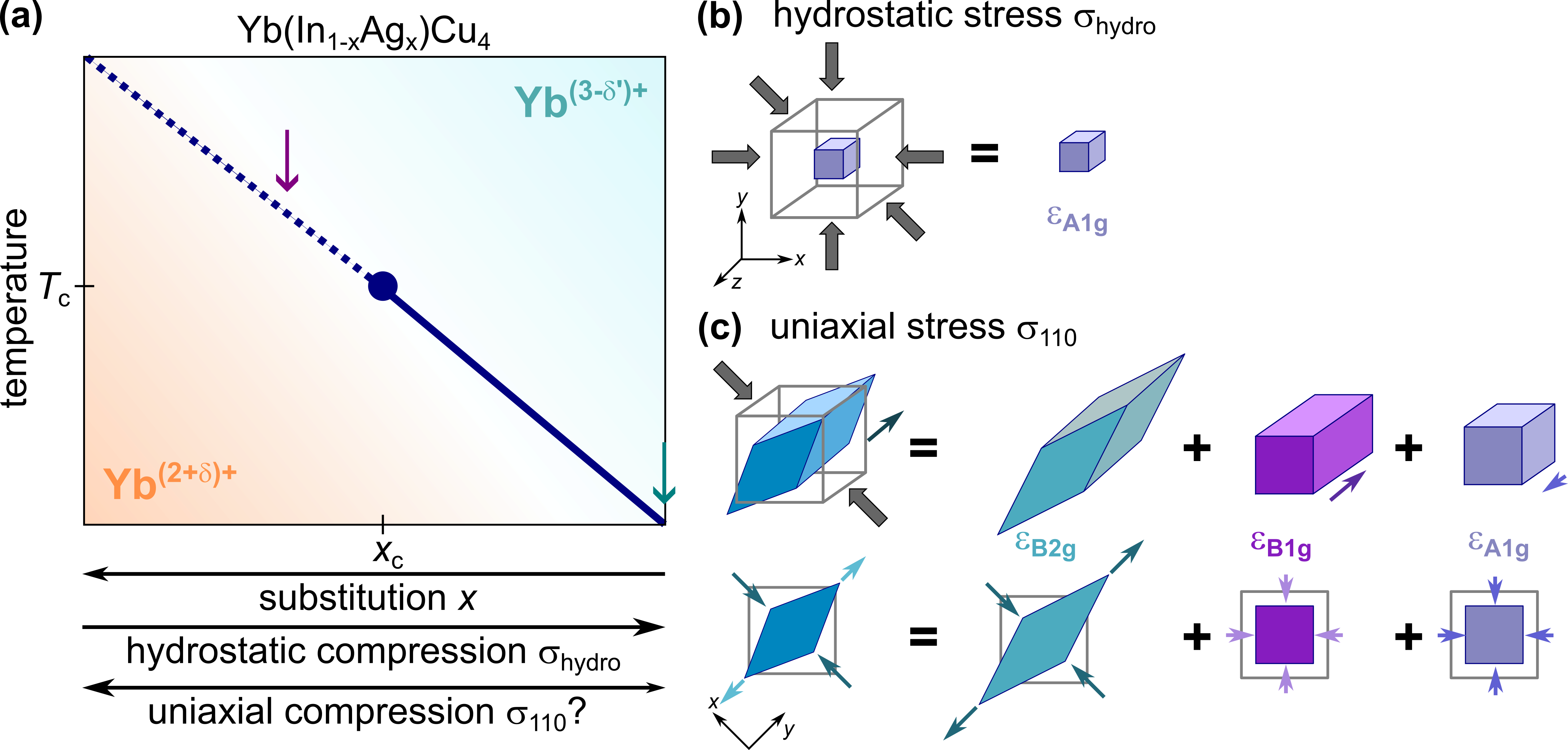}
    \caption{(a) Schematic phase diagram for the valence transition from trivalent to divalent Yb in Yb(In$_{1-x}$Ag$_x$)Cu$_4$ as a function of temperature and external control parameters, such as substitution level, $x$, or hydrostatic stress, $\sigma_\textrm{hydro}$. At low temperatures, the transition is first order (solid line). The first-order transition ends in a second-order critical endpoint, which is located at $(x_c,\,T_c)\,\approx\,(0.1,\,75\,\textnormal{K})$ \cite{Ocker25}. Beyond the critical point, a broad crossover regime (dashed line) is observed. The focus of the present work is on two members of the family with $x\,=\,0$ and $x_\textrm{nom}\,=\,$0.2, whose positions in the general phase diagram are marked by the teal and purple arrows, under uniaxial stress, $\sigma_{[110]}$. (b,c) Schematic illustration of the stress-induced strains in a cubic lattice for hydrostatic stress (b) and uniaxial stress (c). In each panel, the gray cube indicates the original cubic lattice and the bold gray arrows indicate the directions of applied stress. (b) Hydrostatic compression induces symmetry-preserving $\varepsilon_{A1g}$ strains. (c) In contrast, applying uniaxial compression along the diagonal cubic axis [1\,1\,0] breaks the lattice symmetry. Due to the crystal's Poisson ratio, the induced longitudinal and transverse strains are different in magnitude along these three directions, as indicated by the different color of the small arrows. The induced strains can be described by a superposition of irreducible strains of the point group of the crystal lattice: the fully antisymmetric $B_{2g}$ and $B_{1g}$ strains and a fully symmetric $A_{1g}$ strain (for details, see text).}
    \label{fig:schematic}
\end{figure*}

To experimentally establish the widespread relevance of critical elasticity in correlated-electron systems, it is essential to identify additional material realizations where the properties of the critical endpoints are experimentally accessible. Theoretically, the emergence of critical elasticity is governed by symmetry-allowed couplings between the electronic order parameter, $\Phi$, and an appropriate lattice strain component, $\varepsilon_{ij}$ \cite{Zacharias12,Zacharias13,Zacharias15}. Consequently, symmetry analysis provides a powerful framework to guide the search for new candidate systems. In this context, it is noted that the valence transition in rare-earth compounds exhibits striking similarities to the Mott transition, suggesting it as a promising reference case. 

The valence transition, at which the valence of a rare-earth ion changes, can be typically controlled either by temperature or by external parameters, such as chemical substitution or pressure. When the transition is induced at low temperatures, it is first order. This first-order transition line terminates in a finite-temperature second-order critical endpoint, above which only a crossover between the two valence states exists (see Fig.\,\ref{fig:schematic}\,(a)). Since the valence is a scalar order parameter $\Phi$ (like the electronic order parameter of the Mott transition), it can be expected that the electronic critical endpoint is governed by an emergent Ising symmetry. In this case, the symmetry of $\Phi$ allows a linear coupling to symmetry-conserving volumetric strains, $\varepsilon_{A1g}$ (see Fig.\,\ref{fig:schematic}\,(b)), of the form $\Phi\,\cdot\,\epsilon_{ij}$ in the free energy \cite{Zacharias12,Zacharias13}. This condition gives rise to critical elasticity, where the purely electronic critical endpoint is preempted by an isostructural solid-solid endpoint, characterized by a change in volume.

Indeed, the valence transition can in many systems be controlled by hydrostatic stress, $\sigma_\textrm{hydro}$, \cite{Chandran92,Wolf23} consistent with the picture that the transition is primarily driven by volumetric strains. However, given that the valence transition is believed to originate from the interaction between the $f$ electrons and the conduction electrons \cite{Sarrao98}, which can in principle be anisotropic, it is possible that the transition can also be controlled by symmetry-breaking anisotropic strains \cite{Zherlitsyn99}, $\varepsilon_{B2g}$ and $\varepsilon_{B1g}$ (see Fig.\,\ref{fig:schematic}\,(c)). The latter are induced by applying uniaxial stress.

In the present work, we report on the phase diagram of the valence transition under uniaxial stress in the series of cubic Yb(In$_{1-x}$Ag$_x$)Cu$_4$. The pure compound undergoes a first-order valence transition \cite{Sarrao98} as a function of temperature at $T_V\,\approx\,42\,$K, at which the valence changes from the smaller Yb$^{2.9+}$ to the larger Yb$^{2.7+}$ upon cooling \cite{Sarrao96}. Correspondingly, the transition is accompanied by an increase of volume of 0.5\,\%. Upon substituting In with Ag \cite{Sarrao96,Aruga94,Kojima92}, the first-order transition line moves to higher temperatures and terminates in a second-order critical endpoint, above which only a crossover between the two valence states exists. Recent systematic studies of substituted samples \cite{Ocker25} revealed that the critical endpoint is located close to $(x_c,\,T_c)\,\approx\,(0.1,\,75\,\textnormal{K})$. Since hydrostatic compression favors the smaller trivalent state, it generally suppresses the transition temperature $T_V$ and the crossover temperature $T_V^\prime$ (see Fig.\,\ref{fig:schematic}\,(a)). In early work on these compounds, two unusual aspects of the valence transition in YbInCu$_4$ were noted. First, it was pointed out that the small change of the volume is insufficient to explain the large change of the Kondo temperature observed across the valence transition \cite{Cornelius97}. Second, ultrasonic investigations revealed step-like anomalies in the transverse elastic moduli at $T_V$ and $T_V^\prime$, not typical for an isostructural transition \cite{Kindler94,Zherlitsyn99}. The latter result motivates our studies to explicitly investigate the influence of symmetry-breaking strains on the valence transition.

In our experiments, we find that uniaxial stress along the cubic [1\,1\,0] axis is less effective in tuning the valence transition and crossover temperature, $T_V$ and $T_V^\prime$, in Yb(In$_{1-x}$Ag$_x$)Cu$_4$ compared to hydrostatic stress \cite{Sarrao98,Svechkarev99,Matsumoto92,Kojima95,Ocker25}. Using an approach similar to the one developed in Ref.\,\citen{Ikeda18} for an iron-based superconductor, we use this data to disentangle the effects of symmetry-breaking and symmetry-conserving strains on $T_V$ and $T_V^\prime$. The key result of this analysis is that the transition is essentially only tuned by the symmetry-preserving $\varepsilon_{A1g}$ strains, which are much smaller per stress unit in a uniaxial-stress experiment. Our results therefore support the notion that the valence transition is only controlled by changes in the volume and therefore its critical endpoint should be characterized by critical elasticity \cite{Cowley76,Zacharias15}.

\section{Methods}

The single crystals of Yb(In$_{1-x}$Ag$_x$)Cu$_4$ were grown following the procedure described in Ref.\,\citen{Ocker25} with an initial melt composition of 1-1.76-5. The crystals are from batch number MO118 and MO112 for $x_\textrm{nom}$ = 0 and $x_\textrm{nom}$ = 0.2, respectively. Further detailed information on the studied samples, their Ag-concentration and their ambient-stress properties are given in Table \ref{tab:samples}. The ambient-stress transition and crossover temperatures, $T_V$ and $T_V^\prime$, were determined through measurements of the magnetization in a Quantum Design MPMS. The EDX value for the substituted sample was determined for the exact piece of crystal studied under uniaxial pressure.

\begin{table*}[htpb]
\centering
\begin{tabular}{|c|c|c|c|c|c|c|c|}
\hline
$x_\textrm{nom}$ & $x_\textrm{EDX}$ & $T_V$ or & length ($\mu$m)  & width ($\mu$m) & thickness ($\mu$m)\\
& & $T_V^\prime$ (K) & & & \\\hline 0 & 0 & $44.4\,\pm\,0.5$ & $732$ & $81$ & $136$ \\\hline
$0.2$ & $0.126\,\pm\,0.01$ & $87.0\,\pm\,0.5$ & $501$ & $102$ & $125$\\\hline
\end{tabular}
\caption{Summary of samples investigated in the present study. Following the description of the sample characterization in Ref.\,\cite{Ocker25}, we denote the nominal $x$ value and the $x$ value determined by EDX. We also note the transition temperature $T_V$ or the crossover temperature $T_V^\prime$ at ambient stress, determined from a piece of the crystal used in the present work. Finally, dimensions of necked samples, used in the present study, which were cut and measured with a Xenon Plasma Focused Ion Beam (PFIB), are provided.}
\label{tab:samples}
\end{table*}

To study the response to uniaxial stress, we used \textit{in situ}-tunable piezo-driven uniaxial stress cells, similar in design to the one reported in Ref.\,\citen{Barber19}. To this end, the single crystals were polished using a Xenon Plasma Focused Ion Beam (PFIB) into a dumbbell shape with a small neck, which will experience the stress, and larger tabs, which remain essentially unstrained (see Ref.\,\citen{Noad23} and inset of Fig.\,\ref{fig:datax0}\,(a)). The dimensions of the neck are given in Table \ref{tab:samples}, with the long axis being the crystallographic [1\,1\,0] axis, i.e., the axis of applied force. Throughout this work, we denote compressive stresses and strains with a negative sign. 

In order to track $T_V$ and $T_V^\prime$ as a function of $\sigma_{[110]}$, we measure the $T$- and $\sigma_{[110]}$-dependent Young's modulus, $E_{[110]}\,=\,$d$\sigma_{[110]}$/d$\varepsilon_{[110]}$, with $\varepsilon_{[110]}$ being the induced strain along the stress direction. The methods to measure $E_{[110]}$ in piezo-driven uniaxial stress cells are described in Refs. \citen{Noad23} and \citen{ONeil24}. In both methods, the extraction of absolute values of $E_{[110]}$ currently relies on an independent calibration at zero strain/stress, e.g., through ultrasonic measurements of the components of the elastic tensor, $C_{ij}$. Thus, the data presented in this paper is scaled such that it matches Young's modulus obtained from the ultrasonic data of samples with similar composition, reported in Ref.\,\citen{Zherlitsyn99}. Whereas this data can be used to track $T_V$ and $T_V^\prime$, caution should be taken when considering the absolute values of $E_{[110]}$, reported in this work.

\section{Results}

\begin{figure}
    \centering
    \includegraphics[width = \columnwidth]{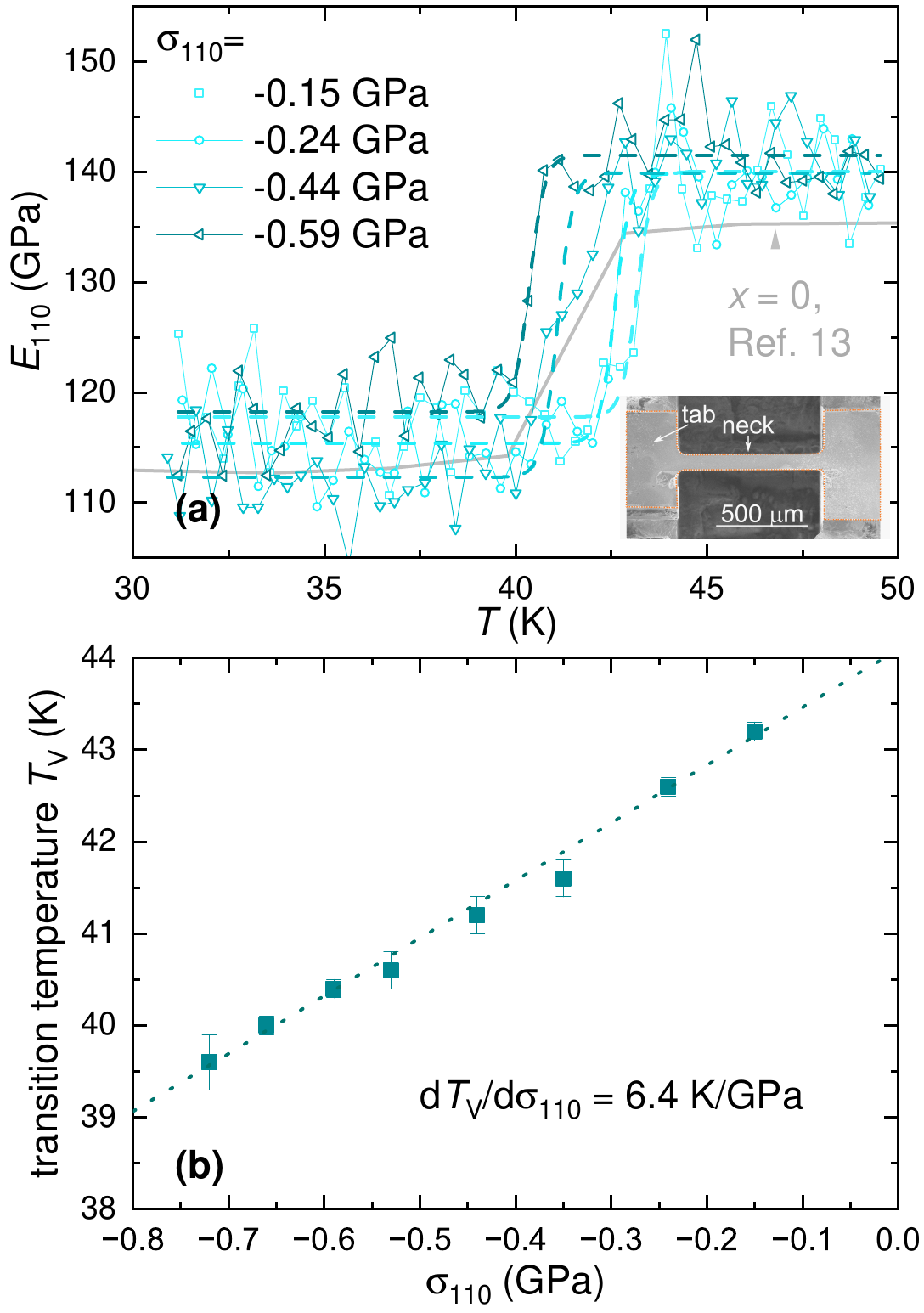}
    \caption{(a) Young's modulus, $E_{[110]}$, of YbInCu$_4$ as a function temperature, $T$, under different uniaxial stress, $\sigma_{[110]}$ (negative sign denotes compression). The dashed lines represent fits to the data with a broadened step function to extract the valence transition temperature, $T_V$. The gray line represents the Young's modulus calculated from ultrasonic data from Ref.\,\cite{Zherlitsyn99}. The inset shows a SEM image of the PFIB-cut sample used for this study. (b) $T_V$ as a function of $\sigma_{[110]}$. The dotted line represents a linear fit to the experimental data.}
    \label{fig:datax0}
\end{figure}

We first discuss data on a pure sample of YbInCu$_4$. Our pre-characterization at ambient stress of the sample used in our study revealed that a first-order transition occurs at $T_V\,\approx\,(44.4\,\pm\,0.5)$\,K. Early ultrasonic measurements \cite{Zherlitsyn99} on a sample of YbInCu$_4$, grown using a slightly different method and with slightly different $T_V$, suggest a step-like change of $E_{[110]}$ at $T_V$ (see gray line in Fig.\,\ref{fig:datax0}\,(a)), consistent with the notion of a first-order transition. As shown in Fig.\,\ref{fig:datax0}\,(a), where we display $E_{[110]}$ vs. $T$ at different negative $\sigma_{[110]}$, we find that the step-like anomaly at $T_V$ shifts to lower temperatures with increasing compression. In order to extract $T_V$, we fit the experimental data by a broadened step function (see dashed lines) and assign the midpoint of the step to $T_V$. The resulting phase diagram of $T_V$ vs. $\sigma_{[110]}$ is shown in Fig.\,\ref{fig:datax0}\,(b). The data is well described by a linear behavior with a slope of d$T_V$/d$\sigma_{[110]}\,=\,(6.4\,\pm\,0.5)$K/GPa.

\begin{figure}
    \centering
    \includegraphics[width = \columnwidth]{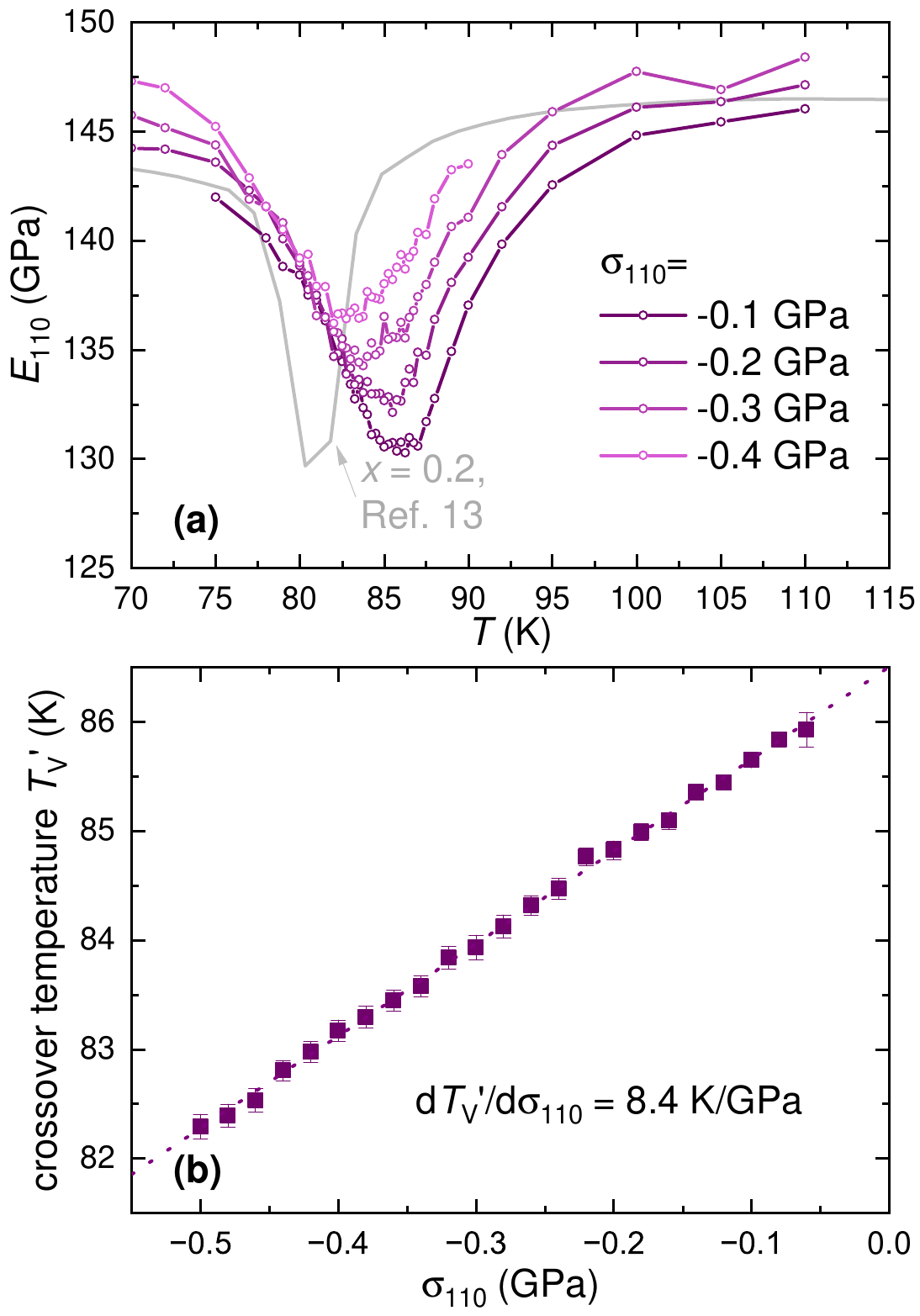}
    \caption{(a) Young's modulus, $E_{[110]}$, of Yb(In$_{1-x}$Ag$_x$)Cu$_4$ ($x_\textrm{nom}\,=\,$0.2) as a function temperature, $T$, under different uniaxial stress, $\sigma_{[110]}$ (negative sign denotes compression). The crossover temperature, $T_V^\prime$, was determined from the minimum of the $E_{[110]}(T)$ curves. The gray line represents the Young's modulus calculated from ultrasonic data from Ref.\,\cite{Zherlitsyn99}. (b) $T_V^\prime$ as a function of $\sigma_{[110]}$. The dotted line represents a linear fit to the experimental data.}
    \label{fig:datax0p2}
\end{figure}

Next we discuss the results for a sample of Yb(In$_{1-x}$Ag$_x$)Cu$_4$ with $x_\textrm{nom}\,=\,$0.2. At ambient stress, this sample undergoes a valence crossover at a characteristic temperature, $T_V^\prime\,=\,(87.0\,\pm\,0.5)$\,K. This crossover manifests itself in a broad minimum in the $T$-dependent $E_{[110]}$, as evident from the earlier ultrasound results \cite{Zherlitsyn99} at ambient stress as well as our data at finite $\sigma_{[110]}$ (see Fig.\,\ref{fig:datax0p2}\,(a)). Similarly to the previous results on the unsubstituted compound, $T_V^\prime$ decreases with increasing compression. We determine $T_V^\prime$ as the minimum in $E_{[110]}(T)$ and plot $T_V^\prime$ as a function of $\sigma_{[110]}$ in Fig.\,\ref{fig:datax0p2}\,(b). The stress dependence is well described by a linear behavior with a slope of d$T_V^\prime$/d$\sigma_{[110]}\,=\,(8.4\,\pm\,0.5)$K/GPa.

\section{Discussion}

The key experimental result of our work is that $T_V$ and $T_V^\prime$ are suppressed linearly with applied $\sigma_{[110]}$. This result may be surprising in light of the sizable antisymmetric $B_{1g}$ and $B_{2g}$ strains in our uniaxial-stress experiment. In general, since the transition temperature, $T_V$, and crossover temperature, $T_V^\prime$, have to be invariant under the symmetry operations of the point group of the crystal (here, among others, the $C_4$ rotation of the cubic lattice), symmetry-breaking strains can only affect the transition and crossover temperatures in even powers, i.e., $\partial T_V/\partial \varepsilon_{B1g}\,=\,\partial T_V/\partial \varepsilon_{B2g}\,=\,0$ and $\partial T_V^\prime/\partial \varepsilon_{B1g}\,=\,\partial T_V^\prime/\partial \varepsilon_{B2g}\,=\,0$. However, symmetry-conserving $\varepsilon_{A1g}$ strains can change the ordering temperature linearly. Thus, we assign the linear $T_V(\sigma_{[110]})$ and $T_V^\prime(\sigma_{[110]})$ behavior, observed in this work, to the response to $A_{1g}$ strains. 

To support this conclusion on a quantitative level, we compare our results to those obtained under hydrostatic stress. To compare the data, we need to explicitly calculate the magnitude of irreducible strains induced in hydrostatic and uniaxial stress experiments (see Figs.\,\ref{fig:schematic}\,(b) and (c)). To this end, we use the elastic compliance tensor $S_{ij}$ for a cubic lattice \cite{Mouhat14}, which is defined in the basis of the crystallographic axes and reprinted in the Appendix Eq.\,\ref{eq:compliance_matrix}. 

In a hydrostatic stress experiment (see Fig.\,\ref{fig:schematic}\,(b)), $\sigma_{11}\,=\,\sigma_{22}\,=\,\sigma_{33}\,=\sigma_\textrm{hydro}$ are finite and all other components zero. Correspondingly, this stress causes all three cubic directions to shrink equally ($\varepsilon_{11}\,=\,\varepsilon_{22}\,=\,\varepsilon_{33}$) and the deformation can therefore be described by a fully symmetric $A_{1g}$ strain only. Using the definition $\varepsilon_{A1g}\,=\,\frac{1}{3}(\varepsilon_{11}+\varepsilon_{22}+\varepsilon_{33})$ is then given by

\begin{eqnarray}
    \varepsilon_\textrm{A1g}\,&=&\,\frac{1}{(C_{11}+2C_{12})} \sigma_\textrm{hydro}. \label{eq:A1ghydr}
\end{eqnarray}

When uniaxial stress is applied along the crystallographic [1\,1\,0] direction, the lattice symmetry is broken. While the crystal is compressed along the [1\,1\,0], it expands along the [1\,-1\,0] and [0\,0\,1] directions (see small arrows in Fig.\,\ref{fig:schematic}\,(c)). To describe the deformations, we need to consider a rotated coordinate system in which the $x$ and $y$ axes are aligned along [1\,1\,0] and [1\,-1\,0], respectively. The rotated compliance matrix, $S^\prime_{ij}$, is given in Appendix Eq.\,\ref{eq:compliance_matrix}. In this frame, the only component of the stress tensor that is finite is $\sigma_{11}\,=\,\sigma_{[110]}$. Correspondingly, the strains read as $\varepsilon_{11}\,=\,S^\prime_{11} \sigma_{[110]}$ (along the [1\,1\,0] direction), $\varepsilon_{22}\,=\,S^\prime_{12} \sigma_{[110]}$ (along the [1\,-1\,0] direction) and $\varepsilon_{33}\,=\,S^\prime_{13} \sigma_{[110]}$ (along the [0\,0\,1] direction). Thus, the crystal strain is different along the $x$, $y$ and $z$ direction. Therefore, in total three irreducible strains are needed to describe the induced deformation (see Fig.\,\ref{fig:schematic}\,(c)). This includes the fully antisymmetric in-plane shear distortion, $\varepsilon_{B2g}\,=\,\frac{1}{2}(\varepsilon_{11}-\varepsilon_{22})$, the tetragonal out-of-plane distortion, $\varepsilon_{B1g}\,=\,\frac{1}{3} (\varepsilon_{33}-(\varepsilon_{11}+\varepsilon_{22}))$ as well as the volumetric $\varepsilon_{A1g}$, defined above. Figure\,\ref{fig:schematic}(c) illustrates the induced irreducible strains, shown from a three-dimensional perspective (top) and as a projection onto the $x$–$y$ plane (bottom). Inserting the explicit expressions for $S^\prime_{ij}$ yields

\begin{eqnarray}
    \varepsilon_\textrm{B2g}\,&=&\,\frac{1}{4 C_{44}}\sigma_{[110]},\\
    \varepsilon_\textrm{B1g}\,&=&-\,\frac{C_{11}+C_{12}}{3(C_{11}^2+C_{11}C_{12}-2C_{12}^2)}\sigma_{[110]},\\
    \varepsilon_\textrm{A1g}\,&=&\,\frac{1}{3} \frac{1}{(C_{11}+2C_{12})} \sigma_{[110]} \label{eq:A1guni}.
    \label{eq:euni}
\end{eqnarray}

Direct comparison of Eqs.\ref{eq:A1ghydr} and \ref{eq:A1guni} shows that the amount of $A_{1g}$ strain induced in a uniaxial-stress experiment is only a third of the one induced in a hydrostatic-stress experiment. Thus, if the stress dependence of $T_V$ and $T_V^\prime$ is solely attributed to the $A_{1g}$ response, then d$T_V$/d$\sigma_\textrm{hydro}\,=\,3\,$d$T_V$/d$\sigma_{[110]}$ and d$T_V^\prime$/d$\sigma_\textrm{hydro}\,=\,3\,$d$T_V^\prime$/d$\sigma_{[110]}$. For the pure compound ($x\,=\,0$), our results of d$T_V$/d$\sigma_{[110]}\,=\,(6.4\,\pm\,0.5)$K/GPa match the expectations of a purely $A_{1g}$-driven response, since d$T_V$/d$\sigma_\textrm{hydro}\,=\,(18\,\pm\,2)$K/GPa, where the error bar reflects the spread of reported values in different studies \cite{Sarrao98,Svechkarev99,Matsumoto92,Kojima95}. Similarly, for the Ag-substituted $x_\textrm{nom}\,=\,$0.2 sample, we find d$T_V^\prime$/d$\sigma_{[110]}\,=\,(8.4\,\pm\,0.5)$K/GPa, which is very close to one third of the  d$T_V^\prime$/d$\sigma_\textrm{hydro}\,=\,25\,$K/GPa, recently reported in Ref.\,\citen{Ocker25}.

The analysis above suggests that the valence transition in the Yb(In$_{1-x}$Ag$_x$)Cu$_4$ family is essentially unaffected by antisymmetric $B_{1g}$ and $B_{2g}$ strains. Overall, the finding that the valence transition is mostly tuned by $A_{1g}$ strains is consistent with the picture that the valence transition is an isostructural phase transition. The critical endpoint of such a transition is predicted to show critical elasticity \cite{Cowley76}, which should experimentally manifest itself in a critically-diverging compressibility $\kappa\,=\,3/(C_{11}+2C_{12})$ (or in other words, a vanishing bulk modulus $B\,=\,\kappa^{-1}$). Since the experimentally studied $E_{[110]}$ in the present work involves a combination of $C_{11}$, $C_{12}$, and $C_{44}$ (see Eq.,\ref{eq:Youngsmodulus}), a precise determination of the bulk modulus near the critical endpoint would require directional-dependent measurements to disentangle these contributions. Such an investigation lies beyond the scope of the present study. Nevertheless, for such an analysis to be meaningful, it remains to be understood why the transverse elastic constants $C_{11} - C_{12}$ and $C_{44}$ exhibit step-like anomalies at the valence transition, despite these being symmetry-forbidden in case of an isostructural transition. Together with the absence of a measurable response of $T_V$ and $T_V^\prime$ to antisymmetric $\varepsilon_{B2g}$ and $\varepsilon_{B1g}$ strains, we speculate that these anomalies in the transverse modes are of extrinsic origin \cite{wolfprivate}. A pronounced sample-to-sample variation in the jump sizes of $C_{11} - C_{12}$ and $C_{44}$ (cf.\,Refs.\,\citen{Zherlitsyn99} and \citen{Kindler94}) at both the transition and crossover supports this scenario. Further studies on the role of disorder and its coupling to elastic degrees of freedom near the critical endpoint \cite{Gati16, Gati18} will be crucial to resolve these questions \cite{wolfprivate}.

In a broader context, the valence transition discussed here represents another instance of a transition that is highly sensitive to symmetry-preserving strains, such as the orbital-selective Mott transition in iron-based superconductors \cite{Wiecki20} or possibly the correlated Mott phases in twisted bilayer graphene \cite{Ma25}. Even though such systems are generally less sensitive to symmetry-breaking strains, uniaxial stress can still be used to control the transition due to the non-negligible symmetric $A_{1g}$ strain that is induced in the experiment. The controllability by uniaxial stress offers the perspective to study the properties of these transitions (such as the valence transition) by surface-sensitive measurements \cite{Jo24,Fedchenko25}, which cannot be conducted in the constraining environment of a hydrostatic stress cell.

\section{Conclusion}

In this work, we studied the temperature-uniaxial stress phase diagram of the valence transition in Yb(In$_{1-x}$Ag$_x$)Cu$_4$. To this end, we studied a sample with $x\,=\,0$, which undergoes a first-order valence transition as a function of temperature, and a sample with $x_\textrm{nom}\,=\,$0.2, for which a valence crossover occurs as a function of temperature. We observe a linear suppression of the valence transition temperature, $T_V$, and the crossover temperature, $T_V^\prime$, with uniaxial stress applied along the crystallographic [1\,1\,0] direction, $\sigma_{[110]}$. Based on a quantitative calculation of the stress-induced strains, we show that d$T_V$/d$\sigma_{[110]}$ and d$T_V^\prime$/d$\sigma_{[110]}$ can be fully accounted for by the response of $T_V$ and $T_V^\prime$ to symmetry-conserving $A_{1g}$ strains, as deduced from hydrostatic stress experiments. As a result, $T_V$ and $T_V^\prime$ are essentially insensitive to symmetry-breaking $B_{1g}$ and $B_{2g}$ strains. The findings of this study are consistent with the picture that the valence transition is solely driven by volumetric effects. As a result, the valence transition in Yb(In$_{1-x}$Ag$_x$)Cu$_4$ fulfills the symmetry conditions \cite{Zacharias12,Zacharias13,Zacharias15} for which the occurrence of critical elasticity has been predicted.

\section{Acknowledgments} We acknowledge useful discussions with Bernd Wolf and Michael Lang. In addition, we gratefully acknowledge funding through the Deutsche Forschungsgemeinschaft (DFG, German Research Foundation) through the TRR 288—422213477 (projects A03 and A13). C.I.O acknowledges the support of a St Leonards scholarship from the University of St. Andrews. Financial support by the Max Planck Society is gratefully acknowledged.

\clearpage
\onecolumngrid
\section{Appendix}

\subsection{Rotation of the compliance matrix}

The compliance matrix $S_{ij}=(C_{ij})^{-1}$ (i.e., the inverse of the elastic constant matrix $C_{ij}$) of a cubic crystal in Voigt notation is given by

\begin{equation}
\begin{pmatrix}
\displaystyle\frac{C_{11} + C_{12}}{C_{11}^2 + C_{11}C_{12} - 2C_{12}^2} & 
\displaystyle -\frac{C_{12}}{C_{11}^2 + C_{11}C_{12} - 2C_{12}^2} & 
\displaystyle -\frac{C_{12}}{C_{11}^2 + C_{11}C_{12} - 2C_{12}^2} & 
0 & 0 & 0 \\
\displaystyle -\frac{C_{12}}{C_{11}^2 + C_{11}C_{12} - 2C_{12}^2} & 
\displaystyle\frac{C_{11} + C_{12}}{C_{11}^2 + C_{11}C_{12} - 2C_{12}^2} & 
\displaystyle -\frac{C_{12}}{C_{11}^2 + C_{11}C_{12} - 2C_{12}^2} & 
0 & 0 & 0 \\
\displaystyle -\frac{C_{12}}{C_{11}^2 + C_{11}C_{12} - 2C_{12}^2} & 
\displaystyle -\frac{C_{12}}{C_{11}^2 + C_{11}C_{12} - 2C_{12}^2} & 
\displaystyle\frac{C_{11} + C_{12}}{C_{11}^2 + C_{11}C_{12} - 2C_{12}^2} & 
0 & 0 & 0 \\
0 & 0 & 0 & \displaystyle\frac{1}{C_{44}} & 0 & 0 \\
0 & 0 & 0 & 0 & \displaystyle\frac{1}{C_{44}} & 0 \\
0 & 0 & 0 & 0 & 0 & \displaystyle\frac{1}{C_{44}} \\
\end{pmatrix}.
\label{eq:compliance_matrix}
\end{equation}

To calculate the induced strains when applying stress along the [1\,1\,0] crystallographic direction, the compliance matrix needs to be expressed in terms of a basis where the two in-plane directions are oriented along [1\,1\,0] and [1\,-1\,0]. To obtain the compliance matrix in this basis, $S^\prime_{ij}$, we rotate $S_{ij}$ by 45$^\circ$ using $t_{\varepsilon} S t_{\sigma}^{-1}$, with 

\begin{align}
t_{\varepsilon} &= 
\begin{pmatrix}
\frac{1}{2} & \frac{1}{2} & 0 & 0 & 0 & \frac{1}{2} \\
\frac{1}{2} & \frac{1}{2} & 0 & 0 & 0 & -\frac{1}{2} \\
0 & 0 & 1 & 0 & 0 & 0 \\
0 & 0 & 0 & \frac{1}{\sqrt{2}} & -\frac{1}{\sqrt{2}} & 0 \\
0 & 0 & 0 & \frac{1}{\sqrt{2}} & \frac{1}{\sqrt{2}} & 0 \\
-1 & 1 & 0 & 0 & 0 & 0
\end{pmatrix}, \\
t_{\sigma}^{-1} &= 
\begin{pmatrix}
\frac{1}{2} & \frac{1}{2} & 0 & 0 & 0 & -1 \\
\frac{1}{2} & \frac{1}{2} & 0 & 0 & 0 & 1 \\
0 & 0 & 1 & 0 & 0 & 0 \\
0 & 0 & 0 & \frac{1}{\sqrt{2}} & \frac{1}{\sqrt{2}} & 0 \\
0 & 0 & 0 & -\frac{1}{\sqrt{2}} & \frac{1}{\sqrt{2}} & 0 \\
\frac{1}{2} & -\frac{1}{2} & 0 & 0 & 0 & 0
\end{pmatrix}.
\end{align}

The result for $S^\prime_{ij}$ reads as

\begin{equation}
\resizebox{0.9\textwidth}{!}{$
\begin{pmatrix}
\displaystyle\frac{C_{11}^2 + C_{11}C_{12} - 2C_{12}^2 + 2C_{11}C_{44}}{4 C_{11}^2 C_{44} + 4 C_{11} C_{12} C_{44} - 8 C_{12}^2 C_{44}} &
\displaystyle -\frac{C_{11}^2 + C_{11}C_{12} - 2C_{12}^2 - 2C_{11}C_{44}}{4 C_{11}^2 C_{44} + 4 C_{11} C_{12} C_{44} - 8 C_{12}^2 C_{44}} &
\displaystyle -\frac{C_{12}}{C_{11}^2 + C_{11}C_{12} - 2C_{12}^2} & 
0 & 0 & 0 \\
\displaystyle -\frac{C_{11}^2 + C_{11}C_{12} - 2C_{12}^2 - 2C_{11}C_{44}}{4 C_{11}^2 C_{44} + 4 C_{11} C_{12} C_{44} - 8 C_{12}^2 C_{44}} &
\displaystyle\frac{C_{11}^2 + C_{11}C_{12} - 2C_{12}^2 + 2C_{11}C_{44}}{4 C_{11}^2 C_{44} + 4 C_{11} C_{12} C_{44} - 8 C_{12}^2 C_{44}} &
\displaystyle -\frac{C_{12}}{C_{11}^2 + C_{11}C_{12} - 2C_{12}^2} & 
0 & 0 & 0 \\
\displaystyle -\frac{C_{12}}{C_{11}^2 + C_{11}C_{12} - 2C_{12}^2} &
\displaystyle -\frac{C_{12}}{C_{11}^2 + C_{11}C_{12} - 2C_{12}^2} &
\displaystyle\frac{C_{11} + C_{12}}{C_{11}^2 + C_{11}C_{12} - 2C_{12}^2} &
0 & 0 & 0 \\
0 & 0 & 0 & \displaystyle\frac{1}{C_{44}} & 0 & 0 \\
0 & 0 & 0 & 0 & \displaystyle\frac{1}{C_{44}} & 0 \\
0 & 0 & 0 & 0 & 0 & \displaystyle\frac{2}{C_{11} - C_{12}} \\
\end{pmatrix}
$}.
\label{eq:rotated_compliance_matrix}
\end{equation}

Correspondingly, the Young's modulus, $E_{[110]}$, is given by

\begin{equation}
    E_{[110]}=(S^\prime_{11})^{-1} = \frac{4 C_{11}^2 C_{44} + 4 C_{11} C_{12} C_{44} - 8 C_{12}^2 C_{44}}{C_{11}^2 + C_{11}C_{12} - 2C_{12}^2 + 2C_{11}C_{44}}
    \label{eq:Youngsmodulus}
\end{equation}

\end{document}